\documentclass[prl,aps,amssymb,twocolumn,showpacs,floatfix]{revtex4}

\usepackage{graphicx}
\usepackage{dcolumn}
\usepackage{bm}

\begin{document}

\preprint{}

\title{Low temperature terahertz spectroscopy of \emph{n}-InSb through a magnetic field driven metal-insulator transition}

\author{X. P. A. Gao, J. Y. Sohn, and S. A. Crooker}

\affiliation{National High Magnetic Field Laboratory, Los Alamos, NM 87545}

\date{\today}
\begin{abstract}
We use fiber-coupled photoconductive emitters and detectors to perform terahertz (THz) spectroscopy
of lightly-doped \emph{n}-InSb directly in the cryogenic (1.5 K) bore of a high-field superconducting
magnet. We measure transmission spectra from 0.1-1.1 THz as the sample is driven through a
metal-insulator transition (MIT) by applied magnetic field. In the low-field metallic state, the
data directly reveal the plasma edge and magneto-plasmon modes. With increasing field, a
surprisingly broad band (0.3-0.8 THz) of low transmission appears at the onset of the MIT. This band
subsequently collapses and evolves into the sharp 1\emph{s}$\rightarrow$2$p^-$ transition of
electrons `frozen' onto isolated donors in the insulating state.
\end{abstract}
 \maketitle
Metal-to-insulator transitions (MITs) in condensed matter have fascinated researchers for decades
\cite{Mott}. MITs have been studied as a function of doping, disorder, pressure and magnetic field
($B$), and two general pictures have emerged \cite{Belitz}. The first concerns Anderson's notion of
disorder-induced localization \cite{PWA}. Alternatively, Mott's picture relates the MIT to the
overlap between electron wavefunctions in a uniform ionic lattice. The inevitable presence of
disorder in real samples usually entangles these two mechanisms, and their effects are hard to
separate. Owing to its continuous tunability, the $B$-induced MITs in lightly-doped semiconductors
(\emph{e.g.} InSb, HgCdTe, GaAs) are of particular interest, for which many mechanisms have been
proposed including magnetic freezeout, Anderson localization, a Mott transition, and even Wigner
crystallization \cite{Keyes, Goldman, Shayegan, Mani}. The more recent observation of an impurity-shifted
cyclotron resonance on the \emph{metallic} side of this MIT in \emph{n}-HgCdTe and \emph{n}-InSb
suggests that $B$ freezes mobile electrons onto donor impurities within a metallic impurity band
\cite{Goldman, Shayegan, Mani}.

Complementing traditional dc transport methods, high frequency spectroscopy of MITs elucidate the
\emph{dynamics} and \emph{excitations} involved. For example, the doping-tuned MIT in NbSi alloys
reveals $\omega/T$ scaling in the ac conductivity from 0.1-1 THz \cite{Lee}, which points to the
quantum-critical nature of this MIT. For the low-density metals realized in lightly-doped
semiconductors, all the relevant frequency (energy) scales generally lie in the THz (meV) range.
Consider Te-doped \emph{n}-InSb, with $n=5 \times 10^{14}$/cm$^3$: The Fermi energy is 1.6 meV
(0.39 THz) and the plasma frequency $\omega_p$=$(ne^2/\epsilon \epsilon_0 m^*)^{1/2} = 2\pi\times$0.42 THz
($\epsilon$$\cong$16 and $m^*$=0.014$m_e$).  The Te donor binding energy is $\sim$2 meV at 1~T \cite{Kaplan}, the
cyclotron frequency $\omega_c$=$eB/m^* \equiv 2\pi\times$2~THz/T, and a MIT is induced at a critical magnetic field
$B_{MIT}$$\sim$1 T \cite{Mani}. It is clearly desirable, therefore, to probe
this $B$-induced MIT using broadband THz spectroscopy spanning these relevant energies.

While bridging the frequency gap between microwave cavity methods and far-infrared (FIR) optical
spectroscopies, standard time-domain THz methods (using free-space lasers, micropositioning stages,
and large THz optics) are not generally suited to the narrow and physically remote low-temperature bores of
high-field solenoids. We therefore developed miniature fiber-coupled emitters and detectors for time-domain THz
spectroscopy directly in the cryogenic bores of superconducting magnets \cite{Crooker}.

Using these recently-developed techniques, we report on the THz transmission spectra through a
series of lightly Te-doped bulk \emph{n}-InSb samples as they are field-tuned through $B_{MIT}$
($n$=2-11$\times$10$^{14}$/cm$^3$). At low $B$, we directly observe the plasma edge and coupled
magneto-plasmon modes of the low-density metal. Most strikingly, as $B$ is tuned through $B_{MIT}$,
we observe an abrupt drop in the THz transmission over a surprisingly broad frequency band (0.3-0.8
THz), which rapidly narrows and evolves into the sharp 1\emph{s}$\rightarrow$2$p^-$ transition (at
$\sim$0.3 THz) characteristic of electrons `frozen' onto Te donors in the insulating state. These spectroscopic terahertz
data therefore provide detailed insight into low-energy electron dynamics through a continuously-tunable MIT.

The experiment (Fig. 1a) is essentially time-domain THz spectroscopy using photoconductive
antennas \cite{Katz}, but with the THz emitter and detector antennas located directly in the
1.5-300~K insert of a high-field, 17~T superconducting solenoid magnet. The antennas are
fiber-coupled, and ultrafast optical pulses from a Ti:Sapphire laser (800 nm) are used to drive
the emitter and gate the detector. Material dispersion in the fibers requires pre-chirping
(stretching) the ultrafast pulses so that they arrive at the antennas, after 20m of fiber, with
$\leq$200~fs pulsewidth (see Ref. \cite{Crooker} for details).  The samples are mounted between the
THz emitter and detector, on a rotating stage that moves the sample in and out of the THz beam
path (for background spectra at each $B$, $T$).

\begin{figure}[tbp]
\includegraphics[width=.43\textwidth]{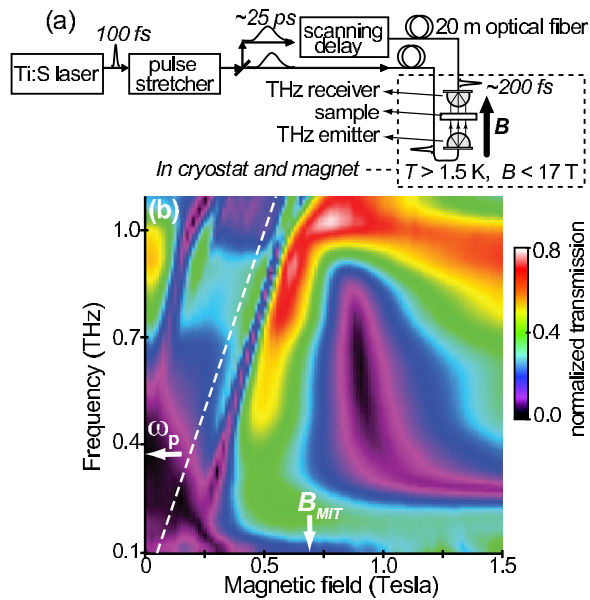}
\caption{(Color online) (a) THz spectroscopy with fiber-coupled antennas. (b) THz transmission spectra through
bulk (0.8 mm thick) Te-doped $n$-InSb vs. $B$ at 1.7 K ($n$=2.1$\times$10$^{14}$/cm$^3$). Black
regions indicate low transmission, from strong absorption or reflection by the sample. Dotted line
represents the one-electron cyclotron frequency, $eB/m^*$.} \label{fig1}
\end{figure}

Figure 1b shows the normalized THz transmission spectra through a 0.8 mm thick sample of
\emph{n}-InSb with $n$=2.1$\times 10^{14}$/cm$^3$, at 1.7 K. Power spectra are recorded vertically,
with frequency increasing from 0.1 to 1.1 THz along the y-axis. $B$ increases from 0 to 1.5 T along
the x-axis. Zero (or low) transmission is represented by a black color, implying strong absorption
or reflection of the THz radiation by the sample. We focus primarily on these black regions, which
indicate the presence of cyclotron resonance (CR), impurity transitions, or magneto-plasma
reflectivity edges. For reference, the CR frequency within a one-electron model (2 THz/T) is plotted as a
dashed white line. At $B$=0 the sample is metallic, and careful inspection reveals no transmission
below 0.38 THz, indicating the total reflection of light below the plasma frequency $\omega_p$ of
this low-density metal. In small $B$, the cyclotron motion of free electrons couples to plasma
oscillations and the plasma edge splits into two magneto-plasmon modes \cite{Palik}: $\omega_\pm =
[(4 \omega_p^2 + \omega_c^2)^{1/2} \pm \omega_c]/2$. These modes are directly observed in the
spectroscopic data. The $\omega_-$ branch is revealed by the border of the low transmission region
at the bottom left, and the $\omega_+$ branch is observed as the dark line with positive slope originating at $\omega_p$.  Lastly, a transition line with slope close to the CR (but shifted to higher $B$) appears for $B\gtrsim0.25$ T (to be discussed). These low-field THz data therefore directly reveal how a free-electron metal evolves into a system of coupled magnetoplasmon modes.

\begin{figure}[tbp]
\includegraphics[width=.45\textwidth]{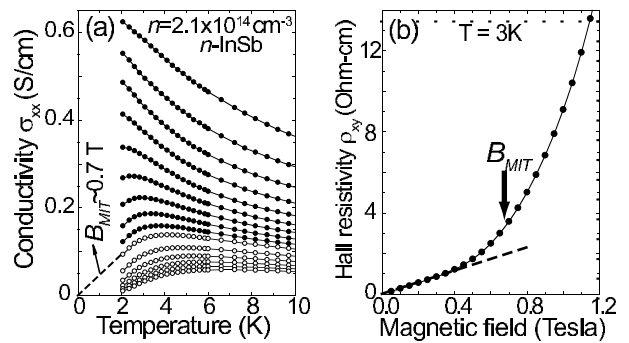}
\caption{(a) The dc conductivity $\sigma_{xx}$ vs. T of $n$-InSb at $B$=.25, .3, .35, .4, .45, .5,
.55, .6, .65, .7, .8, .9, 1.0, 1.1 and 1.2 T (top to bottom). $B_{MIT} \sim 0.7$ T, as given by the
field where $\sigma_{xx}$ extrapolates linearly to zero as $T\rightarrow 0$ (dotted line). (b) The
Hall resistivity $\rho_{xy}$ vs. $B$ at T=3K. Near $B_{MIT}$, $\rho_{xy}$ begins diverging,
indicating electron freeze-out.} \label{fig2}
\end{figure}

The most striking feature in Fig. 1b, however, is the sudden onset
at $B$$\sim$0.8 T of low THz transmission over a wide frequency band from 0.3-0.8 THz.  This region
narrows quickly with increasing $B$, collapsing towards the known 1\emph{s}$\rightarrow$2$p^-$
transition energy ($\sim$0.3 THz) of electrons bound to Te donors \cite{Larsen, Kaplan}. We
associate this marked feature with the onset of the field-induced MIT, and indeed we confirm that $B_{MIT}$=0.7~T in this sample by conventional dc transport. Fig. 2a shows the
temperature-dependent dc conductivity $\sigma_{xx}(T)$ from $B$=0.2-1.2~T (the 2K electron mobility
$\mu=7.7\times 10^4$ cm$^2$/Vs, suggesting a mean scattering time $\tau$=0.6~ps). In the low-field
metallic state $\sigma_{xx}$ tends to finite values as $T\rightarrow 0$, while in the high-field
insulating state $\sigma_{xx} \rightarrow 0$ as $T\rightarrow 0$. As conventionally defined by the
field at which $\sigma_{xx}$ extrapolates linearly to zero as $T\rightarrow 0$ \cite{Shayegan,
Mani}, we find that $B_{MIT}$=0.7~T, in agreement with previous transport studies of
similarly-doped $n$-InSb. Also in agreement with prior work \cite{Shayegan}, the Hall resistivity
$\rho_{xy}$, which is inversely proportional to the free electron density, begins to diverge near
$B_{MIT}$ (see Fig. 2b). These transport data are characteristic of the MIT in \emph{n}-InSb
induced by magnetic freeze-out of the mobile electrons.

The abrupt disappearance of transmitted radiation from 300-800 GHz therefore occurs at magnetic
fields just above the value of $B_{MIT}$ that is conventionally defined by dc transport. The THz data suggest this
phenomenon may be generic -- similar behavior is observed in \emph{all} the $n$-InSb samples
($n$=2.1, 3.4, 5, 6, and 11$\times 10^{14}$/cm$^3$).  Fig. 3a shows THz transmission through the
higher-doped $n=6 \times 10^{14}$/cm$^3$ sample at 1.5~K.  Again, a broad region of low
transmission appears just above $B_{MIT}$ (=1.1~T at this higher density). This region again
narrows as $B > B_{MIT}$, evolving continuously into the sharp 1\emph{s}$\rightarrow$2$p^-$ donor
transition at $\sim$0.3 THz \cite{Larsen, McCombe, Kaplan}. As expected for a
1\emph{s}$\rightarrow$2$p^-$ transition \cite{Kaplan}, we confirm that this absorption shifts to
slightly higher energy (0.4 THz) as $B \rightarrow 15$~T, and disappears for $T$$>$10K (due to thermal ionization of donor-bound electrons).

Importantly, \emph{no sign} of a 1\emph{s}$\rightarrow$2$p^-$ donor transition is observed in the THz spectra at
low $B$ when the sample is in the metallic state. Careful FIR magneto-transmission studies at
single frequencies (0.75-2 THz) have suggested that the MIT in
\emph{n}-InSb occurs within a metallic impurity band \cite{Shayegan}. These new spectroscopic THz
measurements suggest that near $B_{MIT}$, the impurity band may broaden significantly, allowing
excitations over a wide range of frequencies. These data are reminiscent of the doping-tuned MIT in phosphorous-doped Si,
where sharp transitions from donor-bound electrons were observed deep in the insulating
(low-doped) state \cite{Thomas}. These transitions broadened as doping increased toward the
critical density, due to donor-donor interactions and the formation of donor clusters. A similar
picture can be applied to the present case of $n$-InSb. In the insulating state ($B \gg B_{MIT}$),
electrons are ``frozen" onto individual donors and sharp 1\emph{s}$\rightarrow$2$p^-$ transitions from
effectively isolated donors are seen. As $B$ decreases towards $B_{MIT}$, the Bohr radii of bound
electrons grows, interactions and wavefunction overlap become significant, and the notion of a
discrete impurity transition becomes ill-defined. This scenario is consistent with the spirit of a
Mott transition.

\begin{figure}[tbp]
\includegraphics[width=.43\textwidth]{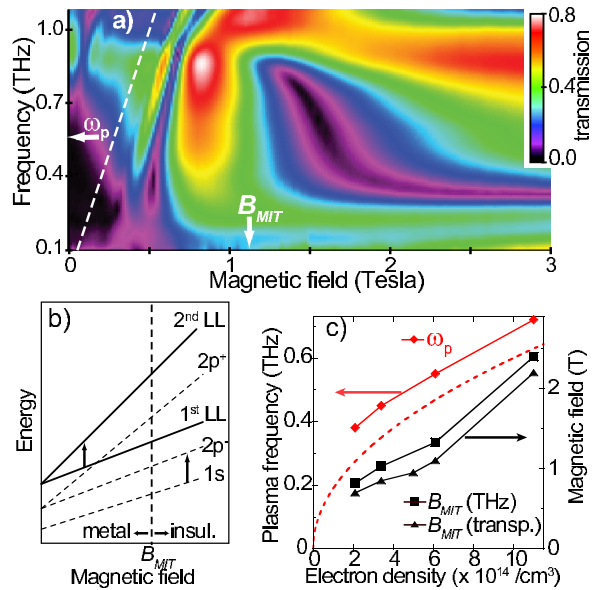}
\caption{(Color online) (a) THz transmission through \emph{n}-InSb vs. $B$ at 1.5 K ($n=6 \times
10^{14}$/cm$^3$). (b) Schematic of the 1st and 2nd Landau levels (solid) and the 1\emph{s}, 2$p^+$,
2$p^-$ donor levels (dashed) vs. $B$. Arrows mark the relevant transitions. (c) Doping dependence
of the measured plasma edge (red diamonds; dashed line is the calculated $\omega_p$), and of
$B_{MIT}$ as measured by transport (triangles) and THz data (squares).} \label{fig3}
\end{figure}

Within a simplified one-electron framework, Fig. 3b sketches the relevant free-electron Landau levels (LL) and donor-bound electron states in \emph{n-}InSb. The 1\emph{s} and 2$p^-$ donor states
follow the 1st LL while the 2$p^+$ donor state tracks the 2nd LL \cite{Kaplan,Palik,McCombe}. Dominant transitions are marked by arrows: i) the (plasmon-shifted) inter-LL CR of free electrons, and ii) the 1\emph{s}$\rightarrow$2$p^-$ transition of electrons frozen onto donors in the insulating state (the 1\emph{s}$\rightarrow$2$p^0$ is forbidden in the Faraday geometry \cite{McCombe}). As a function of electron density, Fig. 3c summarizes the key parameters from the THz magneto-transmission experiments. First, diamonds represent the $B$=0 plasma edge, as measured by the point below which no transmission is observed.  These values agree reasonably well with a calculation of $\omega_p/2\pi$ (dashed line). Secondly, the field at which THz transmission becomes abruptly suppressed (squares) is shown to track $B_{MIT}$ as determined from dc transport (triangles), suggesting their common origin and a possible broadening of an impurity band at $B_{MIT}$.  Note that we cannot distinguish the 1\emph{s}$\rightarrow$2$p^+$ impurity CR \cite{Shayegan, Larsen, McCombe, Kaplan} from the nearby plasma-shifted CR in our
experiment. Further studies may clarify the nature of the additional transition observed in both Figs. 1b and 3a (having slope $\sim eB/m^*$, but shifted to higher $B$/lower energy). Although expected to be weak, a possible transition in these thick samples from the 1st LL$\rightarrow$2$p^+$ is consistent with the observed slope and shift of this line.


These data lend insight into the high-frequency dynamics of electrons undergoing a metal-insulator transition, and highlight the utility of \emph{in situ} low-temperature broadband THz spectroscopy in magnetic
fields using fiber-coupled antennas. Transmission of THz photons through \emph{n}-InSb directly reveals the splitting of the plasma edge into magnetoplasmon modes in the metallic state ($B<B_{MIT}$). In contrast, data in the high-field insulating state ($B \gg B_{MIT}$) directly reveal the sharp 1\emph{s}$\rightarrow$2$p^-$ transition of electrons ``frozen" onto donor impurities. As $B$
decreases toward $B_{MIT}$, the sharp 1\emph{s}$\rightarrow$2$p^-$ transition widens into a broad band with low THz transmission, suggesting the broadening of an impurity band at $B_{MIT}$.  We
thank D. L. Smith, G. S. Boebinger, P. B. Littlewood, B. McCombe, J. Kono, and S. Kos for valuable discussions, and the NHMFL IHRP and Los Alamos LDRD programs for support.


\newpage
\end{document}